\documentclass{article}
\usepackage[utf8]{inputenc}
\usepackage{multirow}
\usepackage{booktabs}
\usepackage{geometry}
\usepackage{filecontents}
\usepackage{caption}
\usepackage{framed}
\usepackage{amsfonts}
\usepackage{amsmath}
\usepackage{mathtools}
\usepackage{graphicx}
\usepackage{courier}
\newcommand{\T}{^\text{T}}

\usepackage{colortbl}
\usepackage{url}
\usepackage{bm}
\usepackage{threeparttable}
\usepackage{array}
\definecolor{linkblue}{RGB}{0,0,200}
\DeclareMathOperator*{\argmax}{argmax}
\usepackage{hyperref}

\usepackage{natbib}
\numberwithin{equation}{section}
\usepackage{lscape}
\usepackage{tikz}

\title{Multi-model mimicry for model selection according to generalised goodness-of-fit criteria}
	
\author{Lachlann McArthur \and Melissa A. Humphries}

\begin{document}

\maketitle

\begin{abstract}
Multi-model mimicry (MMM) is a flexible model selection technique for comparison of multiple, non-nested models on any desired goodness-of-fit criteria. Applicable to any set of candidate models that are 1) able to be fit to observed data, 2) can simulate new sets of data under the models, and 3) have a metric by which a dataset's goodness-of-fit to the model can be calculated, MMM has a much broader range of applicability than many standard model selection techniques. This manuscript highlights the previous literature whilst presenting the theoretical framework underpinning MMM. The scope of applicability is broadened through presentation of generalised criteria for comparison and the effectiveness of the method is demonstrated. Clear instruction for the application of MMM and the classification techniques required for model selection are also included.   
\end{abstract}

At the heart of statistical pursuit is a desire to enumerate actually-occurring phenomena by building models. A statistical approach to scientific enquiry might be said to comprise the following steps:
\begin{enumerate}
	\item Ask a question,
	\item Get relevant data,
	\item Fit models to the data,
	\item Select the best model, and finally
	\item Answer the question.
\end{enumerate}
The fourth step, model selection, is regularly done in frequentist statistics by comparing some goodness-of-fit measure; that is, by selecting the candidate model to which the data best fits, according to some metric. However, in many instances, raw goodness-of-fit measures cannot be directly compared, due to a lack of consistency in the candidate models, data to which they are fit, or goodness-of-fit measure. The limitations of direct comparisons of likelihood-based goodness-of-fit measures, as well as the desirability of using alternative bases for model comparison, are those sought to be overcome by the multi-model mimicry (MMM) framework. A broadly-applicable method for model comparison on the basis of any desired goodness-of-fit measure, MMM is a potential forerunner in new model selection techniques designed to cope with the increasingly-complex analytical methods evolving in the statistical field.

For the first time, this manuscript presents the genesis and statistical theory behind the MMM. Beginning with Wilks' likelihood-ratio test for nested hypotheses, and extending into general pairs of non-nested hypotheses \citep{Williams:1970ab}, sensitivity to model parameters and dependence on likelihood as a measure of goodness-of-fit are highlighted as reasons to consider alternate model selection techniques. A simplified, pairwise model mimicry technique of Wagenmakers \emph{et al.} \citep{Wagenmakers:2004aa} is the foundational next step, before considering a recently-proposed extension to multiple models, the multi-model mimicry (MMM) framework. The theoretical underpinning of the MMM framework are provided, along with clear instructions for the scope of its’ applicability including advantages and limitations of the technique. The efficacy of the MMM approach is then demonstrated, using simulated examples, and new suggestions for both classification and visualisation are suggested.

\section{Likelihood-based goodness-of-fit comparisons and their limitations} \label{sec:liklim}
Traditional goodness-of-fit measures are based around likelihood, which is defined as, for observed data \(\bm{x}\), and model \(\mathcal{M}\) with parameter set \(\theta\),
	\begin{equation*}
	\mathcal{L}_\mathcal{M}(\theta |\bm{x}) = f_{\mathcal{M}}(\bm{x}|\theta),
	\end{equation*}
where \(f_{\mathcal{M}}\) is the probability distribution for data under the model \(\mathcal{M}\). Since likelihood can in most instances be increased by arbitrarily increasing model complexity, more sophisticated goodness-of-fit measures penalise complexity, measured as the number of parameters \(k\) estimated in the model. For example, Akaike's Information Criterion (AIC) is defined \citep{Akaike:1974aa} as
	\begin{equation*}
	\text{AIC}=2k - 2\log(\hat{\mathcal{L}}_\mathcal{M}),
	\end{equation*}
while the Bayesian Information Criterion (BIC) is given \citep{Schwarz:1978aa} by
	\begin{equation*}
	\text{BIC}=k\log(n) - 2\log(\hat{\mathcal{L}}_\mathcal{M}),
	\end{equation*}
	where \(\hat{\mathcal{L}}_\mathcal{M}\) is the likelihood of model \(\hat{\mathcal{M}}\) under maximum likelihood parameters \(\hat{\theta}\), and \(n\) is the number of data points in observed data set \(\bm{x}\). A model with the smallest AIC or BIC is selected. Naturally, BIC penalises complexity more than AIC (unless \(\log{n} < 2\), which should never occur in practice as this implies that there are 7 or fewer observed data points).
	
	AIC- or BIC-based model selection, however, is limited in the situations in which it can be applied. Situations in which AIC or BIC are inappropriate include:
	\begin{itemize}
	    \item where likelihood is intractable, in theory or practice;
	    \item where the structure of the set of models, or data upon which the models are compared, makes AIC or BIC inappropriate; and
	    \item where alternative model selection bases are deemed more appropriate, for philosophical and/or practical reasons.
	\end{itemize}
	Each of these situations will now be discussed in turn.
	
	\subsection{Intractable likelihood}
A likelihood may be impossible to calculate for one or more of the candidate models. Types of data and models for which this might occur are well documented in the literature; examples include networks \citep{Caimo:2015aa, Ratmann:2007aa, Ratmann:2009aa}, complicated time series \citep{Breto:2009aa, Jasra:2014aa}, and hidden Markov models \citep{Yildirim:2015aa}. In differentiating between these types of models, Approximate Bayesian Computation (ABC) has recently gained popularity, but this technique is sensitive to prior distributions for both the choice of model and for each model's parameters, as well as to choices of summary statistics \citep{Robert:2011aa}. This manuscript presents an alternative manner of differentiating between models, without the selection of prior probabilities in the model space. 

	
\subsection{Model or data structure}
Some model structures are of sufficiently different form to be incomparable using likelihood-based methods like AIC or BIC. For example, suppose some univariate data of sample size \(n\) is to be fit either using a normal distribution, or using kernel density estimates. AIC or BIC require calculation of likelihood, and of the number of parameters. The first model, the normal distribution, has two parameters---the mean and standard deviation. For the kernel density model, the `number of parameters' is somewhat nebulous, as all observed data points are considered for all estimates, but have differing influence. 

The structure of data might also lead to models being incomparable using AIC or BIC. For example, suppose two time series models are to be considered, an ARIMA model, and an ARIMA model applied to differenced data. There is one more data point in the ARIMA model than in the differenced ARIMA model. As such, the likelihood function of the ARIMA model will contain one more term than will the differenced ARIMA model, so will be of a different order of magnitude \citep{Harvey:1980aa}. This makes direct comparison of likelihoods, and likelihood-based goodness-of-fit measures, unavailable. The MMM procedure in this manuscript overcomes this issue using a simulation-based approach.
	
\subsection{Alternative model selection bases}
Finally, likelihood is not always the preferred criterion according to which we wish to select a model. One example of this is in choosing an appropriate distribution to fit to some given data. Since likelihood measures the probability density of the observed data, treated usually as independent observations, given a model, it does not take into consideration whether the data fits the shape of the proposed distribution; that is, likelihood is not designed to differentiate whether a candidate distribution is appropriate given the skew, kurtosis or other moments of the data.

In this instance, it might be desirable to compare candidate distributions on the basis of a distributional goodness-of-fit measure, such as Kolmogorov-Smirnov statistic \citep{Massey:1951aa}, Sk\`ekely and Rizzo's energy statistic \citep{Szekely:2005aa}, or some more rudimentary summary statistic like the number of extreme values in a distribution. Since these statistics' raw values cannot be directly compared between candidate models, a more rigorous framework to compare these values must be considered. The MMM framework in this manuscript is able to address this.
	
A similar motivation may be to compare models on the basis of how reasonable the models' assumptions are. Goodness-of-fit to model assumptions is not a simple binary question; some valid models are more well-founded than others in terms of their assumptions. For example, consider a simple linear regression model. One assumption is normality of the error term, which is usually assessed using a quantile-quantile plot of the residuals. The assumption of normality might be considered reasonable for a number of different quantile-quantile plots, but one might more readily accept some plots than others. This is one demonstration of the fact that the validity of an assumption is on a continuum. Where a model fits on this continuum for some distributional assumption may distinguish some candidate models from others. 

Since fidelity to assumptions justifies the generalisability of a model, a model might be favoured if it satisfies its assumptions better than other candidates. However, it is not immediately obvious how one might compare the fidelity of two or more models to their respective distributional assumptions. For example, if two models assume different error distributions, comparing these assumptions would involve comparing the goodness-of-fit of one set of errors to one distribution, with the goodness-of-fit of another set of errors to a different distribution. Again, test statistics for distributional fit, like the Kolmogorov-Smirnoff or energy statistics \citep{Szekely:2005aa}, are not measured on the same scale for all candidate distributions, and may be more sensitive to some kinds of lack-of-fit than to others. These raw goodness-of-fit statistics thus cannot be directly compared between models. The problem of comparing distributional goodness-of-fit among such models is explored in detail in this manuscript, with the MMM framework able to provide for such comparisons.

\section{Genesis of general goodness-of-fit comparisons} \label{sec:GOFnon-nes}

\subsection{Likelihood-ratio test for nested hypotheses}
The likelihood-ratio statistic (Wilks 1938) is a well-known basis for hypothesis tests comparing two nested models \citep{Wilks:1938aa}. When models are nested, they come from the same parameterised family, so the hypothesis test consists of choosing between two sets of parameters for this family, often denoted \(\theta_A\) and \(\theta_B\). The likelihood-ratio statistic is \[\Lambda = \frac{\mathcal{L}(\theta_A | \bm{x})}{\mathcal{L}(\theta_B | \bm{x})},\] for observed data \(\bm{x}\) and common likelihood function \(\mathcal{L}\). Significance levels for this statistic are easily determined, since for nested models, \[ -2\log (\Lambda) \stackrel{\mathcal{D}}{\to}  \chi^2_{p_B- p_A} \] as sample size \(n\to\infty\), with \(p_B\) the dimensionality of \(\theta_B\) and \(p_A\) the dimensionality of \(\theta_A\) (Wilks 1938) \citep{Wilks:1938aa}. 

This test, however, cannot be undertaken for non-nested hypotheses, since in this case, the likelihoods in the expression for \(\Lambda\) would be from different families of distribution, meaning the distribution of \(\Lambda\) does not necessarily follow Wilks' asymptotic expression.

\subsection{Extension to non-nested hypotheses}\label{sec:CoxRL}
In response to this shortcoming, Cox (1961) \citep{Cox:1961aa} extends the likelihood-ratio test to some cases with non-nested hypotheses. 
Suppose we have some realisations \(\bm{x}\) of random variable \(X\), and seek to compare two hypotheses:
\begin{align*}
H_A: X \sim A(\theta_A),\\
H_B: X \sim B(\theta_B),
\end{align*}
where \(A\) and \(B\) are non-nested distributions, and \(H_A\) is the null hypothesis. Cox' test statistic compares the logarithm of the ratio of likelihoods under each hypothesis to the expected value of the log-ratio under the null hypothesis. Cox' statistic can be written as
\[
T = \log \left(  \frac{\mathcal{L}_{A}(\theta_A | \bm{x})}{\mathcal{L}_{B}(\theta_B | \bm{x})}  \right)   - E_{A(\theta_A)} \left[ \log \left(   
\frac{\mathcal{L}_{A}(\theta_A | \bm{x})}{\mathcal{L}_{B}(\theta_B | \bm{x})}
\right)\right],
\] where \(\mathcal{L}_{A}\) and \(\mathcal{L}_{B}\) are likelihood functions of distributions \(A\) and \(B\) respectively, and \(E_{A(\theta_A)}\) denotes an expectation with respect to distribution \(A\) with parameters \(\theta_A\).

Cox demonstrates that \(T\) is asymptotically normally distributed, with mean zero. He notes that the variance of \(T\), and the expectation term in the expression for \(T\), is difficult to calculate, depending on the distributions \(A\) and \(B\) \citep{Cox:1961aa}. Derivations of these values for a limited number of pairs of non-nested hypotheses have been published (see, e.g., \citep{Cox:1962aa, Jackson:1968aa, Walker:1967aa}).

\subsection{A Monte Carlo approach to test statistic distributions} \label{sec:CoxMC}
Due to the intractability of the asymptotic distribution of Cox' statistic \(T\) for some pairs of non-nested hypotheses, and the fact that \(T\) often converges to its asymptotic distribution slowly (Williams 1970) \citep{Williams:1970ab}, Williams introduces a simulation approach to determining a distribution of the test statistic. Using an equivalent variation on the test statistic,
\[
\lambda = \log \left(  \frac{\mathcal{L}_{A}(\theta_A | \bm{x})}{\mathcal{L}_{B}(\theta_B | \bm{x})}  \right),
\] Williams proposes simulating distributions for \(\lambda\) under both the null and alternative hypotheses, and then drawing a conclusion as to which hypothesis is to be favoured \citep{Williams:1970aa, Williams:1970ab}. More explicitly, his approach consists of the following steps:
\begin{itemize}
	\item Estimate parameters \(\theta_A\) and \(\theta_B\), for null distribution \(A\) and alternative distribution \(B\) respectively, from the \(n\) observations in observed data \(\bm{x}\), and calculate an observed value of \(\lambda\).
	\item Simulate \(R\) datasets of size \(n\) under \(A(\theta_A)\), denoted \(\bm{x}^r_A: r=1,\dots,R\), and \(R\) datasets of size \(n\) under \(B(\theta_B)\), denoted \(\bm{x}^r_B: r=1,\dots,R\).
	\item For each simulation \(r=1, \dots, R\), calculate \[
	\lambda^r_A = \log \left(  \frac{\mathcal{L}_{A}(\theta_A | \bm{x}^r_A)}{\mathcal{L}_{B}(\theta_B | \bm{x}^r_A)}  \right) \text{ and } \lambda^r_B = \log \left(  \frac{\mathcal{L}_{A}(\theta_A | \bm{x}^r_B)}{\mathcal{L}_{B}(\theta_B | \bm{x}^r_B)}  \right).
	\]
	\item Compare the observed value of \(\lambda\) to the distributions of the \(\lambda^r_A\)s and the \(\lambda^r_B\)s: 
	\subitem The null hypothesis, that \(X \sim A(\theta_A)\), is to be preferred if the observed value of \(\lambda\) is more likely under the distribution of \(\lambda^r_A\) than of \(\lambda^r_B\). Otherwise, the alternative hypothesis, that \(X \sim B(\theta_B)\), is to be preferred.
\end{itemize}
If the observed value of \(\lambda\) is unlikely under both distributions, it may be that neither model is preferable, and if the observed value of \(\lambda\) is likely under both models, there is insufficient evidence to prefer one model over another \citep{Williams:1970aa}.

Williams proposes \(R=10\) due to the historical computing limitations, but larger values can now be chosen to give greater confidence in model choice \citep{Williams:1970ab}. 

Williams notes that a limitation of his approach is that the simulated distribution of the test statistic is strongly dependent upon the values of the estimated parameters \(\theta_A\) and \(\theta_B\), and suggests further simulation may alleviate this \citep{Williams:1970ab}. He does not suggest a specific method for doing so. The model mimicry method accounts for this parameter uncertainty using a bootstrap.

\section{The model mimicry method} \label{sec:modelmimicry}
\subsection{The model mimicry method for two models} \label{sec:PBCM}
As noted in Section \ref{sec:CoxMC}, the approach of Williams in non-nested model selection is hindered by the fact that it does not account for parameter uncertainty \citep{Williams:1970ab, Wagenmakers:2004aa}. 

An additional limitation of his approach is that it is built around a likelihood ratio goodness-of-fit measure \citep{Wagenmakers:2004aa}. While this is suitable for a number of model comparison situations, popular measures like the Akaike information criterion (AIC) and the Bayes information criterion (BIC) might be sought to be compared across non-nested models. 

Other goodness-of-fit measures may also be preferred in specific instances. For example, we might seek a model whose distributional assumptions are best justified. A goodness-of-fit measure for multivariate distributions is thus more appropriate here than comparing likelihoods. 

Accounting for these drawbacks of the approach of Williams (1970), Wagenmakers \emph{et al.} (2004) \citep{Wagenmakers:2004aa} present a method for testing hypotheses of non-nested models which both accounts for uncertainty in parameter estimation, and is suited to general goodness-of-fit measures. The method consists of re-framing Williams' approach in terms of a generic goodness-of-fit measure, and adding the additional step of a non-parametric bootstrap prior to each simulation. The non-parametric bootstrap precludes the distribution of goodness-of-fit measures from relying heavily on a particular parameter estimate; instead, for a stable model, a variety of parameters will be used, drawn from the region of the parameter space inhabited by those estimated using the observed data.

Wagenmakers \emph{et al.} call this approach ``model mimicry" \citep{Wagenmakers:2004aa}. This is because the method leads to the selection of models that best replicate the observed data. The specific method they propose is labelled the ``parametric bootstrap cross-fitting method" (`PBCM'), since the act of simulating data under each model is a parametric bootstrap, and each model is fit to both models' simulations. In this manuscript, this is referred to as ``model mimicry'', so as to provide a clearer distinction with the method later introduced for comparing multiple models (``multi-model mimicry'')

\begin{figure}
	\begin{center}
\includegraphics[width=0.8\textwidth]{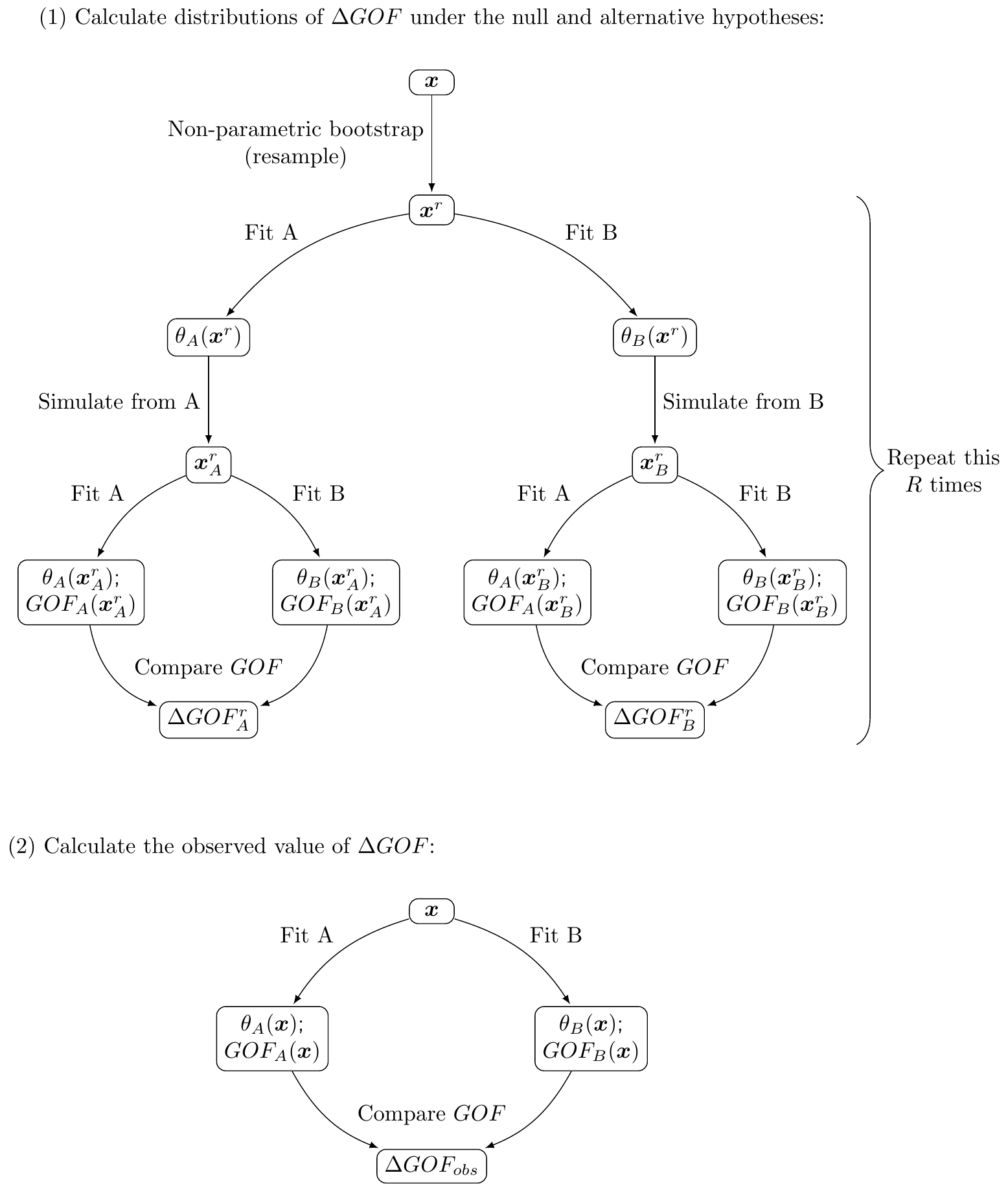}
		\caption[The model mimicry method outlined by Wagenmakers \emph{et al.} (2004)]{The model mimicry outlined by Wagenmakers \emph{et al.} (2004). Simulated distributions of differences in goodness-of-fit under competing models are compared to the difference between the models' goodness-of-fit for the observed data.}
\label{fig:PBCMdiagram}
\end{center}
\end{figure}

The model mimicry approach of Wagenmakers \emph{et al.} is as follows \citep{Wagenmakers:2004aa}:
\begin{enumerate}
	\item Apply the non-parametric bootstrap to the \(n\) observations in observed data \(\bm{x}\); in other words, take a sample of size \(n\) from \(\bm{x}\), sampling with replacement. Denote this \(\bm{x}^r\).
	\item Estimate parameters \(\theta_A(\bm{x}^r)\) and \(\theta_B(\bm{x}^r)\), for null distribution \(A\) and alternative distribution \(B\) respectively, from the \(n\) observations in bootstrap \(\bm{x}^r\).
	\item Simulate a dataset of size \(n\) under \(\theta_A(\bm{x}^r)\), denoted \(\bm{x}^r_A\), and a dataset of size \(n\) under \(\theta_B(\bm{x}^r)\), denoted \(\bm{x}^r_B\).
	\item Fit both models to both sets of simulated data, and calculate goodness-of-fit (`GOF') measures for each of the models' fit. In other words:
	\begin{itemize}
	\item Estimate parameters \(\theta_A(\bm{x}^r_A)\) and \(\theta_B(\bm{x}^r_A)\), for null distribution \(A\) and alternative distribution \(B\) respectively, from the \(n\) observations in \(\bm{x}^r_A\), and calculate GOF measures \(GOF_A(\bm{x}^r_A)\) and \(GOF_B(\bm{x}^r_A)\); and
	\item Estimate parameters \(\theta_A(\bm{x}^r_B)\) and \(\theta_B(\bm{x}^r_B)\), for null distribution \(A\) and alternative distribution \(B\) respectively, from the \(n\) observations in \(\bm{x}^r_B\), and calculate GOF measures \(GOF_A(\bm{x}^r_B)\) and \(GOF_B(\bm{x}^r_B)\).
	\end{itemize}
	\item For the data generated from \(A\), calculate the difference in the goodness-of-fit measures between the two models: \[\Delta GOF^r_A =    GOF_A(\bm{x}^r_A) -GOF_B(\bm{x}^r_A).\] Do the same for the data generated from \(B\): \[\Delta GOF^r_B =    GOF_A(\bm{x}^r_B) -GOF_B(\bm{x}^r_B).\]
	\item Repeat steps 1-5 for \(r = 1, \dots, R\), yielding \(R\) observations from the distribution of \(\Delta GOF_A\) and of \(\Delta GOF_B\).
	\item Meanwhile, fit both models to the observed data \(\bm{x}\), yielding \(\theta_A(\bm{x})\) and \(\theta_B(\bm{x})\). Calculate the goodness-of-fit of each model, \(GOF_A(\bm{x})\) and \(GOF_B(\bm{x})\), and the difference between these: \[\Delta GOF_{obs} = GOF_A(\bm{x})-GOF_B(\bm{x}).\]
		\item Compare the observed value  \(\Delta GOF_{obs}\) to the distributions of \(\Delta GOF_A\) and \(\Delta GOF_B\): 
		\begin{itemize}
	\item The null hypothesis, that \(X \sim A(\theta_A)\), is to be preferred if the observed value \(\Delta GOF_{obs} \) is more likely under the distribution of \(\Delta GOF_A\) than of \(\Delta GOF_B\); otherwise, the alternative hypothesis, that \(X \sim B(\theta_B)\), is to be preferred.
	\item In other words, select model \(A\) if \[
	\frac{f(\Delta GOF_{obs}| A \text{ is true})}{ f(\Delta GOF_{obs}| B \text{ is true}) } \geq 1,
	\] for density function \(f\).
	\end{itemize}
\end{enumerate}
A diagram outlining the model mimicry method can be found in Figure \ref{fig:PBCMdiagram}.

\paragraph{Data from a Cauchy distribution:}
To illustrate this process, a comparison was made between the fit of the normal and Cauchy distributions to data simulated according to a Cauchy distribution. Using the distribution \(\text{Cauchy}(1,5)\), 100 variates were simulated. Model mimicry was applied, with 500 replicates, comparing the normal distribution to the Cauchy distribution. The goodness-of-fit statistic chosen for comparison of fit to the distributions was Sk\`ekely and Rizzo's energy statistic \citep{Szekely:2005aa}; this demonstrates the notion that model mimicry can be applied in a much broader range of situations than likelihood-based approaches.

The energy statistic relies upon the idea that the structure of Euclidean distances between independently-drawn deviates from a given distribution is unique to that distribution. In other words, if two distributions \(\bm{X}\) and \(\bm{Y}\) are different, the expected distance between one point from \(\bm{X}\) and one point from \(\bm{Y}\) should be greater than the mean of: (1) the expected distance between two points from \(\bm{X}\); and (2) the expected distance between two points from \(\bm{Y}\). The energy statistic uses this argument to measure the distance between observed data and a proposed distribution.

\begin{figure}
	\begin{center}
		\includegraphics[width=0.5\textwidth]{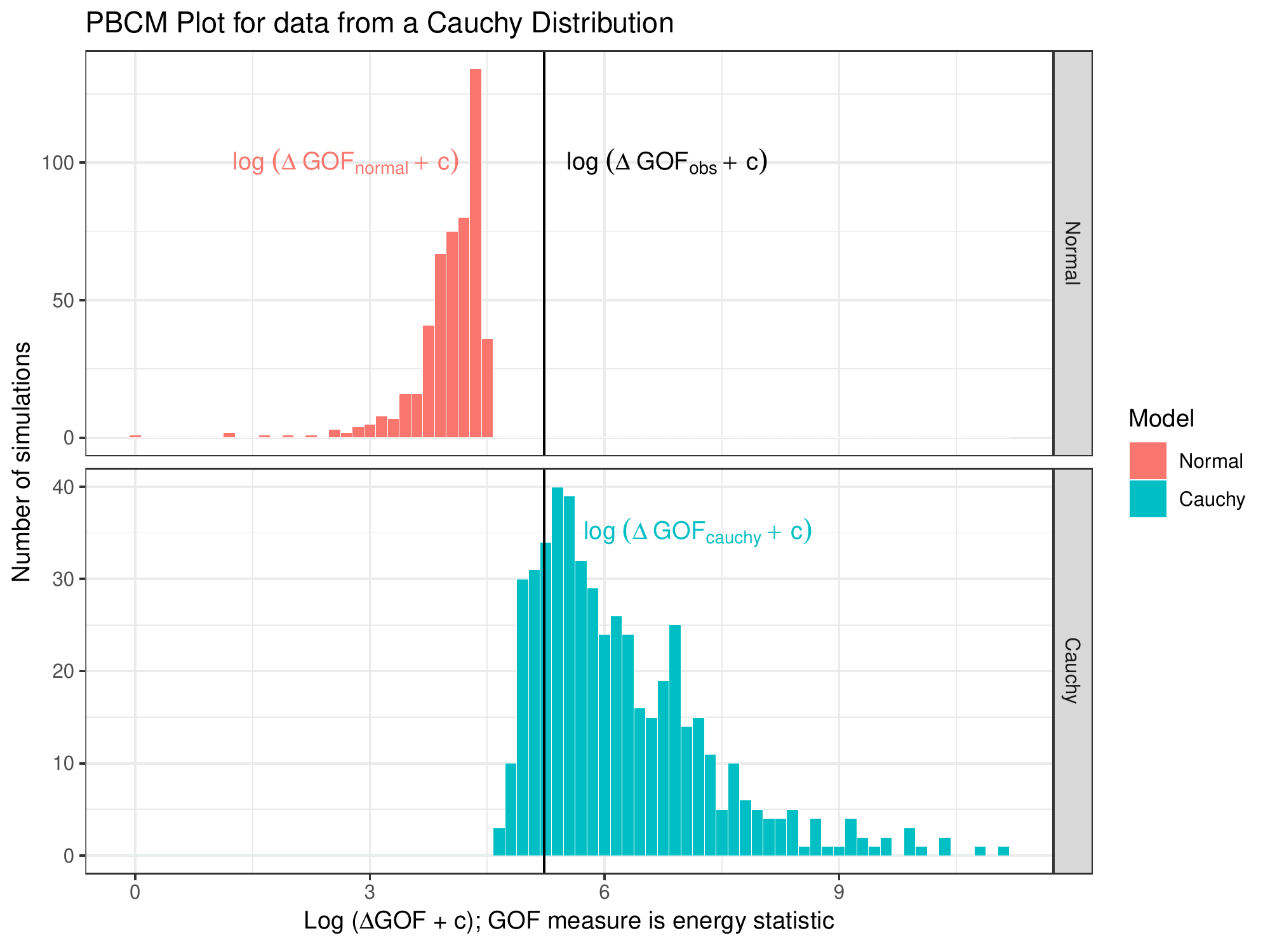} 
		\caption[Example of the PBCM, for simulated Cauchy data]{Plot comparing the fit of the normal (orange) and Cauchy (green) distributions to data simulated from a Cauchy distribution, using the model mimicry method. The black line is the difference in goodness of fit to the observed data and this is nested within the Cauchy distribution meaning the Cauchy distribution appears to be the best fit to the data.}
		\label{fig:PBCMegcauchy}
	\end{center}
\end{figure}

The application of the model mimicry technique yielded 500 observations from a distribution of \(\Delta GOF_{normal}\) and \(\Delta GOF_{Cauchy}\), the difference in energy statistics between the normal and the Cauchy distributions when the true model is assumed to be normal and Cauchy respectively. For a visual representation, a plot of logarithms of the two \(\Delta GOF\) distributions, shifted by a constant \(c\), with a vertical line for \(\log (\Delta GOF_{obs}+c)\), can be found in Figure \ref{fig:PBCMegcauchy}. Natural logarithms needed to be taken due to the very large variation in empirical energy statistics for the Cauchy distribution. The constant is added to ensure all values are positive, since logarithms can only be taken of positive values. 

Logarithms were taken due to large variation in energy statistics for the Cauchy distribution. The observed value, \(\bm{GOF_{obs}}\), is represented by a black line. It is clear from Figure \ref{fig:PBCMegcauchy} that a Cauchy distribution is a better fit to the Cauchy simulated data than a normal distribution, since \[f(\Delta GOF_{obs}| X\sim\text{Cauchy} (x_0, \gamma)) > f(\Delta GOF_{obs}| X\sim N(\mu,\sigma^2))  , \] for model parameters \(\mu, \sigma^2, x_0,\) and \(\gamma\), and true data distribution \(X\). 

As seen in this example, simulated distributions of goodness-of-fit measures provide a useful reference by which to compare the fit of two models. When the observed difference in goodness-of-fit between two models is more likely under one model than another, this model is to be preferred, regardless of whether the raw goodness-of-fit statistic is higher for one model than for another. The model mimicry method, in being able to compare models for which goodness-of-fit statistics are not directly comparable, and being able to compare irregular models like the Cauchy distribution, presents a robust method for complicated model comparisons.

\subsection{Extensions to more than two models} \label{sec:MMPBCM}
Wagenmakers \emph{et al.}'s model mimicry is limited in that it allows for the comparison of only two models. The method of comparing distributions of differences in goodness-of-fit does not easily extend to greater than two models, so when more than two non-nested candidate models are to be compared, another approach should be taken. 

While more than two models can be compared on a pairwise basis, it should be noted that Wagenmakers \emph{et al.}'s model mimicry assumes in each comparison that one of the competing models is true. The decision rule for model selection encouraged by the Wagenmakers \emph{et al.}'s model mimicry, that we should prefer model \(A\) over \(B\) if 
\[
\frac{f(\Delta GOF_{obs}| A \text{ is true})}{ f(\Delta GOF_{obs}| B \text{ is true}) } \geq 1,
\] is less useful when neither model \(A\) nor \(B\) is true. 

For example, if comparing on a pairwise basis three models \(A, B\) and \(C\), for which model \(A\) is true, one of the three comparisons will be undertaken on an incorrect assumption--that between model \(B\) and model \(C\). This, however, will not be a fatal problem if model \(A\) is indeed true; the pairwise applications of the Wagenmakers \emph{et al.}'s model mimicry should reveal model \(A\) is preferable to both models \(B\) and \(C\), rendering the comparison between models \(B\) and \(C\) unnecessary. If one is willing to accept the discomfort of undertaking particular pairwise analyses on flawed assumptions, the pairwise model selection procedure may be appropriate.

If one seeks to avoid this by comparing all models simultaneously, the similar approaches of Allcroft and Glasbey (2003) \citep{Allcroft:2003aa} and Schultheis and Naidu (2014) \citep{Schultheis:2014aa} may be preferred. Allcroft and Glasbey (2003) \citep{Allcroft:2003aa}, one year before Wagenmakers \emph{et al.} published their methodology \citep{Wagenmakers:2004aa}, propose a technique similar to Wagenmakers \emph{et al.}'s model mimicry, but with the capability to compare more than two models. The major point of difference between the Allcroft and Glasbey method and Wagenmakers \emph{et al.}'s model mimicry, is that while \(\Delta GOF=GOF_1-GOF_2\) distributions are simulated in model mimicry, the Allcroft and Glasbey method uses raw \(GOF\) values to simulate multivariate distributions of \([GOF_1, GOF_2, \dots, GOF_M]\) under each of models \(1,2,\dots,M\). The observed value of \([GOF_1, GOF_2, \dots, GOF_M]\) is then compared to the \(M\) simulated distributions to determine which hypothesis is most likely. After simulation, model selection then becomes a classification problem in an \(M\)-dimensional space. 

In contrast to Wagenmakers \emph{et al.}'s model mimicry, the Allcroft and Glasbey method omits the non-parametric bootstrap at each simulation, and does not re-estimate the parameters of each model for each simulation, instead using only the estimated parameters from the observed data throughout the procedure. These omissions are reversed in the work of Schultheis and Naidu (2014) \citep{Schultheis:2014aa}, and their technique will be preferred here to reduce the procedure's sensitivity to parameter estimates. 

\begin{figure}
	\begin{center}
		\includegraphics[width=0.9\textwidth]{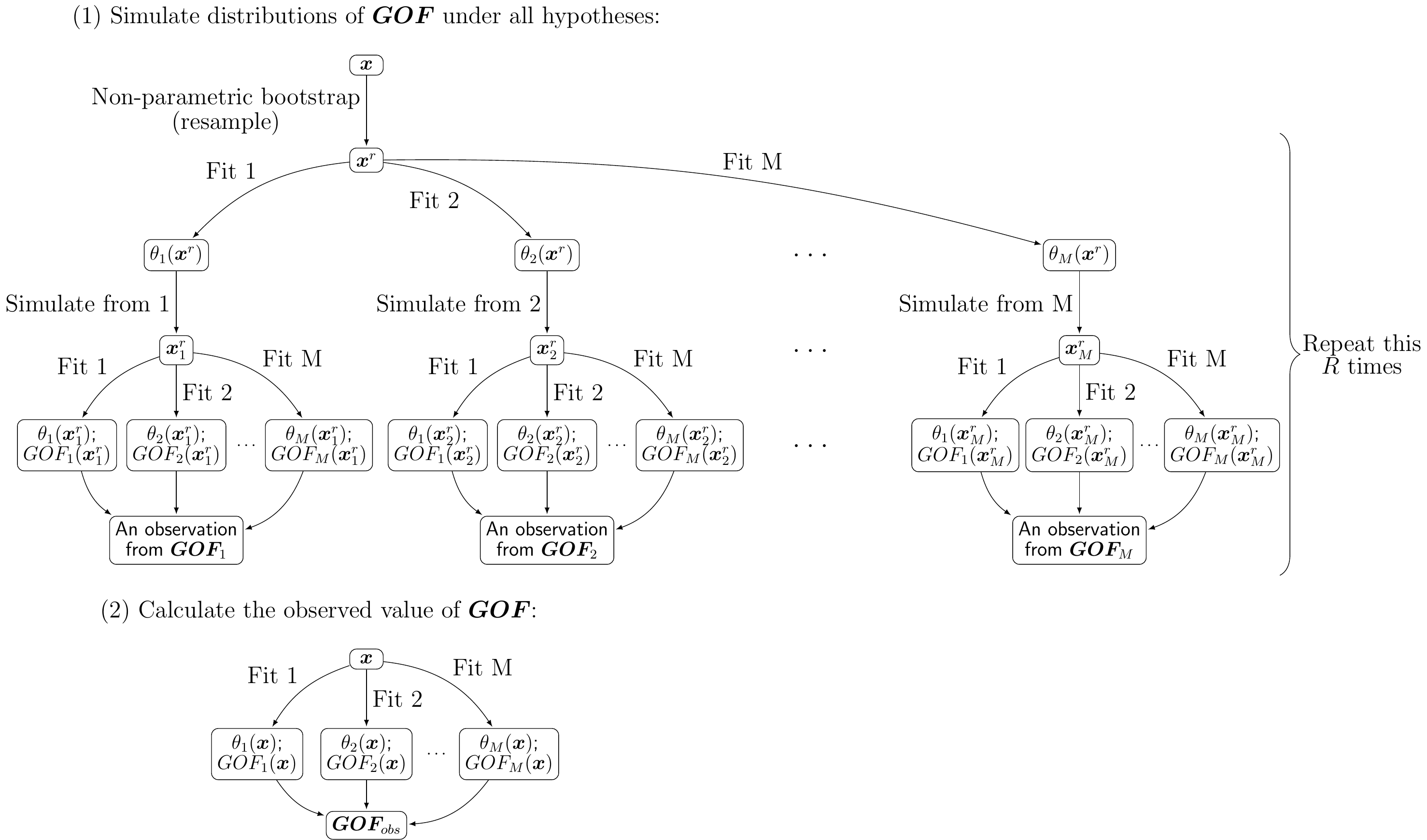}
		\caption[Multi-model mimicry outlined by Schultheis and Naidu (2014)]{Multi-model mimicry outlined by Schultheis and Naidu (2014). Simulated multivariate distributions of goodness-of-fit under competing models are compared to all models' goodness-of-fit to the observed data.}
		\label{fig:PBCMdiagrammulti}
	\end{center}
\end{figure}

The favoured method, of Schultheis and Naidu (2014), is here called ``multi-model mimicry'', is thus as follows \citep{Schultheis:2014aa}:

\begin{enumerate}
	\item Apply the non-parametric bootstrap to the \(n\) observations in observed data \(\bm{x}\); in other words, take a sample of size \(n\) from \(\bm{x}\), sampling with replacement. Denote this \(\bm{x}^r\).
	\item Estimate parameters \(\theta_1(\bm{x}^r),  \theta_2(\bm{x}^r), \dots \theta_M(\bm{x}^r)\), for proposed distributions \(1, 2, \dots, M\) respectively, from the \(n\) observations in bootstrap \(\bm{x}^r\).
	\item Simulate dataset of size \(n\) under each of \(\theta_1(\bm{x}^r),  \theta_2(\bm{x}^r), \dots \theta_M(\bm{x}^r)\), denoted, \(\bm{x}^r_1, \bm{x}^r_2, \dots, \bm{x}^r_M\) respectively.
	\item Fit every model to every set of simulated data, and calculate goodness-of-fit (`GOF') measures for each of the models' fit. In other words:
	\begin{itemize}
		\item Estimate parameters \(\theta_1(\bm{x}^r_1), \theta_2(\bm{x}^r_1), \dots, \theta_M(\bm{x}^r_1)\), for  distributions \(1, 2, \dots, M\) respectively, from the \(n\) observations in \(\bm{x}^r_1\), and calculate GOF measures \(GOF_1(\bm{x}^r_1), GOF_2(\bm{x}^r_1), \dots, GOF_M(\bm{x}^r_1)\);
		\item Estimate parameters \(\theta_1(\bm{x}^r_2), \theta_2(\bm{x}^r_2), \dots, \theta_M(\bm{x}^r_2)\), for  distributions \(1, 2, \dots, M\) respectively, from the \(n\) observations in \(\bm{x}^r_2\), and calculate GOF measures \(GOF_1(\bm{x}^r_2), GOF_2(\bm{x}^r_2), \dots, GOF_M(\bm{x}^r_2)\);
		\begin{center}\(\vdots\)\end{center}
		\item Estimate parameters \(\theta_1(\bm{x}^r_M), \theta_2(\bm{x}^r_M), \dots, \theta_M(\bm{x}^r_M)\), for  distributions \(1, 2, \dots, M\) respectively, from the \(n\) observations in \(\bm{x}^r_M\), and calculate GOF measures \(GOF_1(\bm{x}^r_M), GOF_2(\bm{x}^r_M), \dots, GOF_M(\bm{x}^r_M)\).
	\end{itemize}
	\item Repeat steps 1-4 for \(r = 1, \dots, R\), yielding \(R\) observations from the distributions of
	 \begin{align*}
	 \bm{GOF}_1 &= [GOF_{1|1}, GOF_{2|1}, \dots GOF_{M|1}],\\
	 \bm{GOF}_2 &= [GOF_{1|2}, GOF_{2|2}, \dots GOF_{M|2}],\\
	 &\vdots\\
	 \bm{GOF}_M & =  [GOF_{1|M}, GOF_{2|M}, \dots GOF_{M|M}],
	 \end{align*}
	 where \(GOF_{i|j}\) is the goodness-of-fit to model \(i\) of data produced according to model \(j\).
	\item Meanwhile, fit all models to the observed data \(\bm{x}\), yielding \(\theta_1(\bm{x}), \theta_2(\bm{x}),\) \(\dots,\theta_M(\bm{x}) \). Calculate the goodness-of-fit of each model,	 \[\bm{GOF}_{obs} = [GOF_1(\bm{x}), GOF_2(\bm{x}),\dots, GOF_M(\bm{x})] .\]
	\item Compare the observed value \(\bm{GOF}_{obs} \) to the distributions of \(\bm{GOF}_1\), \(\bm{GOF}_2\), \(\dots\) \(\bm{GOF}_M\): 
	\begin{itemize}
		\item The hypothesis to be preferred is that under which the observed value \(\bm{GOF}_{obs} \) is most likely.
		\item In other words, select the model satisfying \[
		\displaystyle\argmax_{i=1,\dots,M}  f(\bm{GOF}_{obs} | \bm{GOF}_{obs} \sim \bm{GOF}_{i}).
		\]
	\end{itemize}
\end{enumerate}
A diagram outlining the multiple model comparisons method can be found in Figure \ref{fig:PBCMdiagrammulti}.

\subsection{Classifying results from multi-model mimicry} \label{sec:PBCMclass}
In Step 7 of the method adapted here from Schultheis and Naidu \citep{Schultheis:2014aa}, a model is chosen which satisfies \[
		\displaystyle\argmax_{i=1,\dots,M}  f(\bm{GOF}_{obs} | \bm{GOF}_{obs} \sim \bm{GOF}_{i}).
		\]
Unfortunately, the distributions of \(\bm{GOF}_{i}, i=1,2,\dots,M\) are only known through the \(R\) simulated observations of these distributions. The task of choosing the model that maximises the density of \(\bm{GOF}_{obs}\) under that model is, in other words, a supervised classification task, assigning a new point to one of \(M\) sets of observed points. Three popular methods for supervised classification are \begin{itemize}
    \item inspection,
    \item model-based classifiers, and
    \item non-parametric classifiers.
\end{itemize}
The first two such methods are discussed in the paragraphs below. For further discussion of non-parametric classifiers, including \(k\)-nearest neighbour methods, see Hastie, Tibshirani and Friedman \citep{Hastie:2009aa}; these methods are not used here as the decision boundaries in these scenarios are rarely so irregular as to necessitate such an approach.

\paragraph{Inspection}
With just a single data point to classify, it seems natural at first to determine the nearest distribution by inspection. In this context, this would amount to looking at a plot containing \(\bm{GOF_{obs}}\) and all observations of \(\bm{GOF}_{i}, i=1,2,\dots,M\), and determining which set of observations appears closest to \(\bm{GOF_{obs}}\). For simple analyses, there is nothing inherently wrong with this approach; indeed, in their paper introducing model mimicry, Wagenmakers \emph{et al.} use inspection to classify models for their Example 1 \citep{Wagenmakers:2004aa}. However, there are two key flaws with this methodology; that inspection becomes more difficult in a higher-dimensional space, and that inspection may become inefficient for multiple data sets.

In the examples of Wagenmakers\emph{ et al.} \citep{Wagenmakers:2004aa}, two models are differentiated using model mimicry; that is, the version able to deal with a binary model selection. Wagenmakers\emph{ et al.} were thus able to make model selections by inspecting histograms. For multi-model mimicry, multi-dimensional distributions of points are considered, making visualisation of \(\bm{GOF_{obs}}\) within the goodness-of-fit space difficult. Pairwise scatterplots (by dimension) are possible, though information is lost in showing just marginal goodness-of-fit distributions. An alternative is a two-dimensional principal components plot, which is able to represent a much larger proportion of the variation in the data than a two-dimensional marginal plot. This is one visualisation used by Schultheis and Naidu \citep{Schultheis:2014aa}, though the principal components plot is also unlikely to fully convey the higher-dimensional system it represents. This manuscript will thus propose the use of discriminant-based classifiers for multi-model mimicry output.

Another issue with model selection by inspection is that it becomes inefficient when a larger number of data sets are considered. For example, suppose a model is sought to describe multiple potential realisations of data sets. This situation is common in psychological modelling (see, e.g., Wagenmakers \emph{et al.} \citep{Wagenmakers:2004aa}). In this instance, it may be necessary to classify many values of \(\bm{GOF_{obs}}\), with separate sets of distributions of \(\bm{GOF}_{i}, i=1,2,\dots,M\) for each data set. By inspection, this would require considering a large number of visualisations in order to draw conclusions. Numerical classification of \(\bm{GOF_{obs}}\) should then be considered.

\paragraph{Discriminant-based classifiers}
Numerically, we seek the probability that \(\bm{GOF_{obs}}\) belongs to some model \(\mathcal{M}\). In other words, for each \(i=1,2,\dots,M\), we seek \[
P(\mathcal{M}=i \;|\; \bm{GOF_{obs}}).
\]
Using Bayes' theorem, we can rearrange this to 
\begin{align}
    P(\mathcal{M}=i \;|\; \bm{GOF_{obs}})&= \frac{f(\bm{GOF_{obs}} \;|\; \mathcal{M}=i)P(\mathcal{M}=i)}{f(\bm{GOF_{obs}})} \nonumber\\
    &= \frac{f(\bm{GOF_{obs}} \;|\; \mathcal{M}=i)P(\mathcal{M}=i)}{\sum_{j=1}^M f(\bm{GOF_{obs}} \;|\; \mathcal{M}=j)P(\mathcal{M}=j)}. \nonumber
    \intertext{Since there are the same number (denoted earlier \(R\)) of observations for each model, and there is no assumed prior preference for any model, we can treat \(P(\mathcal{M}=i)=P(\mathcal{M}=j) \;\forall\; i,j\):}
    P(\mathcal{M}=i \;|\; \bm{GOF_{obs}})&=\frac{f(\bm{GOF_{obs}} \;|\; \mathcal{M}=i)}{\sum_{j=1}^M f(\bm{GOF_{obs}} \;|\; \mathcal{M}=j)}. \label{eq:pbcmclassifier}
\end{align}
In order to classify \(\bm{GOF_{obs}}\), we thus need to estimate \(f(\bm{GOF_{obs}} \;|\; \mathcal{M}=i)\). We briefly discuss here estimation of the distributions of \(\bm{GOF_{obs}} \;|\; \mathcal{M}=i\) in three manners of increasing complexity (all from Hastie, Tibshirani and Friedman (2009) \citep{Hastie:2009aa}):
\begin{itemize}
    \item Linear discriminant analysis (LDA), which assumes homoscedastic normal distributions;
    \item Quadratic discriminant analysis (QDA), which assumes heteroscedastic normal distributions; and
    \item Mixture discriminant analysis (MDA), which assumes mixtures of heteroscedastic normal distributions.
\end{itemize}
It should be noted that Hastie, Tibshirani and Friedman restrict their version of MDA to the assumption of mixtures of homoscedastic normal distributions (\citep{Hastie:2009aa} at page 440), while this is generalised here to mixtures of heteroscedastic normal distributions for additional flexibility.

\emph{Linear discriminant analysis} (LDA), the simplest of the three methods, is named because it produces linear decision boundaries; in other words, the boundary between the region whose points that would be classified to one model, and the region that would be classified to another, is always linear \citep{Hastie:2009aa}. The LDA model supposes that each distribution can be expressed as
\[
\bm{GOF_{obs}} \;|\; \mathcal{M}=i \;\sim\; N(\bm{\mu_i}, \Sigma),
\] for \(i=1,2,\dots,M\). Note here that \(\Sigma\) is not dependent on \(i\), meaning the variance is assumed to be the same for all models. Calculating the density functions can be done by maximum likelihood; each mean \(\bm{\mu}_i\) is simply estimated to be the sample mean of each \(\bm{GOF}_{i}\), while the variance \(\Sigma\) is estimated to be the weighted average of sample variance matrices \(\hat{\Sigma}_i\) for each \(\bm{GOF}_{i}\). Since in this case, there are the same number of observations of each goodness-of-fit distribution, \[
\hat{\Sigma} = \frac{1}{M}\sum_{i=1}^M \hat{\Sigma}_i,
\] for variance estimate \(\hat{\Sigma}\), and sample variances \(\hat{\Sigma}_i\) estimated by \begin{align}
\hat{\Sigma}_i = \frac{1}{R-1} \sum_{r=1}^{R} (\bm{GOF}_{i,r} - \overline{\bm{GOF}}_{i})(\bm{GOF}_{i,r} - \overline{\bm{GOF}}_{i})\T\label{eq:sampvar},
\end{align} 
for \(\bm{GOF}_{i,r}\) the \(i\)th observation of \(\bm{GOF}_{i}\). Once these parameters have been estimated, Equation \eqref{eq:pbcmclassifier} can be used to select a model; the model that maximises the discriminant in \eqref{eq:pbcmclassifier} can be selected. The higher the value of the discriminant for the chosen model, the greater the confidence one can express in the selection of that model.

\emph{Quadratic discriminant analysis} (QDA) relaxes the constant variance assumption from LDA. This means it is able to respond to differing covariance structures between goodness-of-fit distributions. The example later in this manuscript demonstrates that this may be useful in a multi-model mimicry context; in that instance, the variance of distributions, as represented in Figure \ref{fig:PBCMegpairs}, was not the same for all candidate models. 

The QDA model supposes that each distribution can be expressed as
\[
\bm{GOF_{obs}} \;|\; \mathcal{M}=i \;\sim\; N(\bm{\mu_i}, \Sigma_i),
\] for \(i=1,2,\dots,M\). Note here that \(\Sigma_i\) is dependent on \(i\). Again, density functions are calculated by maximum likelihood, with means for each model estimated by sample mean \(\overline{\bm{GOF}}_{i}\), and sample variances as in Equation \eqref{eq:sampvar} above.

\emph{Mixture discriminant analysis} (MDA) presupposes that \(\bm{GOF_{obs}} \;|\; \mathcal{M}=i\) can be approximated by a Gaussian mixture, which is a weighted sum of normal random variables. Thus MDA allows for more flexible decision boundaries than LDA or QDA, by accounting for the fact that the distributions of goodness-of-fit among candidate models may be multi-modal. The MDA model supposes that each goodness-of-fit distribution can be expressed as
\[
\bm{GOF_{obs}} \;|\; \mathcal{M}=i \;\sim\; \sum_{l=1}^{k_i} \pi_{i,l}  N(\bm{\mu_{i,l}}, \Sigma_{i,l}),
\] for \({k_i}\) components of the \(i\)th mixture model, \(i=1,2,\dots,M\). Note here that the means, variances and even number of components may vary within and among candidate distributions. 

The allowance for the covariance matrices being not identical for all components and for all \(\bm{GOF}_{i}\) distributions is an extension of the discriminant suggested by Hastie, Tibshirani and Friedman (\citep{Hastie:2009aa} at page 440). The additional flexibility provided by this extension is necessary in situations such as that in the example in Figure \ref{fig:PBCMegPCA}, in which the distribution of goodness-of-fit under the log-normal candidate model, unlike the distribution for other candidate models, has a highly irregular covariance structure. Only a more flexible Gaussian mixture is able to capture this irregular covariance structure, which requires multiple components with different covariance matrices.

The parameters of the mixture model, within each group, can be estimated using an Expectation Maximisation (EM) algorithm for Gaussian mixtures \citep{Xu:1996aa}.

Using LDA, QDA or MDA, models can be selected by maximising the discriminant in Equation \eqref{eq:pbcmclassifier}. If the LDA, QDA or MDA assumptions hold, the discriminant in Equation \eqref{eq:pbcmclassifier} represents an estimate of the probability that each model is true, given \(\bm{GOF}_{obs}\), using an uninformative prior \citep{Hastie:2009aa}. This provides for a numerical representation of not just which model should be selected, but also the uncertainty in this choice of model.

\subsection{An example of multi-model mimicry} \label{sec:pbcmegmm}
Using the expansion of the procedure of Schultheis and Naidu (2014) \citep{Schultheis:2014aa}, this section contains an example concerning data simulated from an exponential distribution. Hypotheses that the data come from an exponential distribution, a log-normal distribution and a chi-squared distribution are compared using multi-model mimicry, using Sk\`ekely and Rizzo's energy statistic \citep{Szekely:2005aa} as a measure of goodness-of-fit to each distribution, as in this manuscript's earlier example with regard to pairwise model mimicry. 

Since the three candidate distributions are similar, they should be difficult to distinguish, making this example demonstrative of the power of multi-model mimicry.

First, 100 variates from the distribution Exp\((1)\) were simulated. Multi-model mimicry was then applied to this data, comparing the ability of an exponential distribution, a log-normal distribution and a chi-squared distribution to describe the data, with 500 replicates. This yielded 500 observations from the distributions of \(\bm{GOF_{exp}},\) \(\bm{GOF_{logN}}\) and \(\bm{GOF_{\chi^2}}\), the energy goodness-of-fit of all models to the data simulated under the exponential, log-normal and chi-squared models respectively. Each model was also fit to the observed data, yielding the observed goodness-of-fit of each model to the data, \(\bm{GOF_{obs}}\).

\begin{figure}
	\begin{center}
		\includegraphics[width=0.5\textwidth]{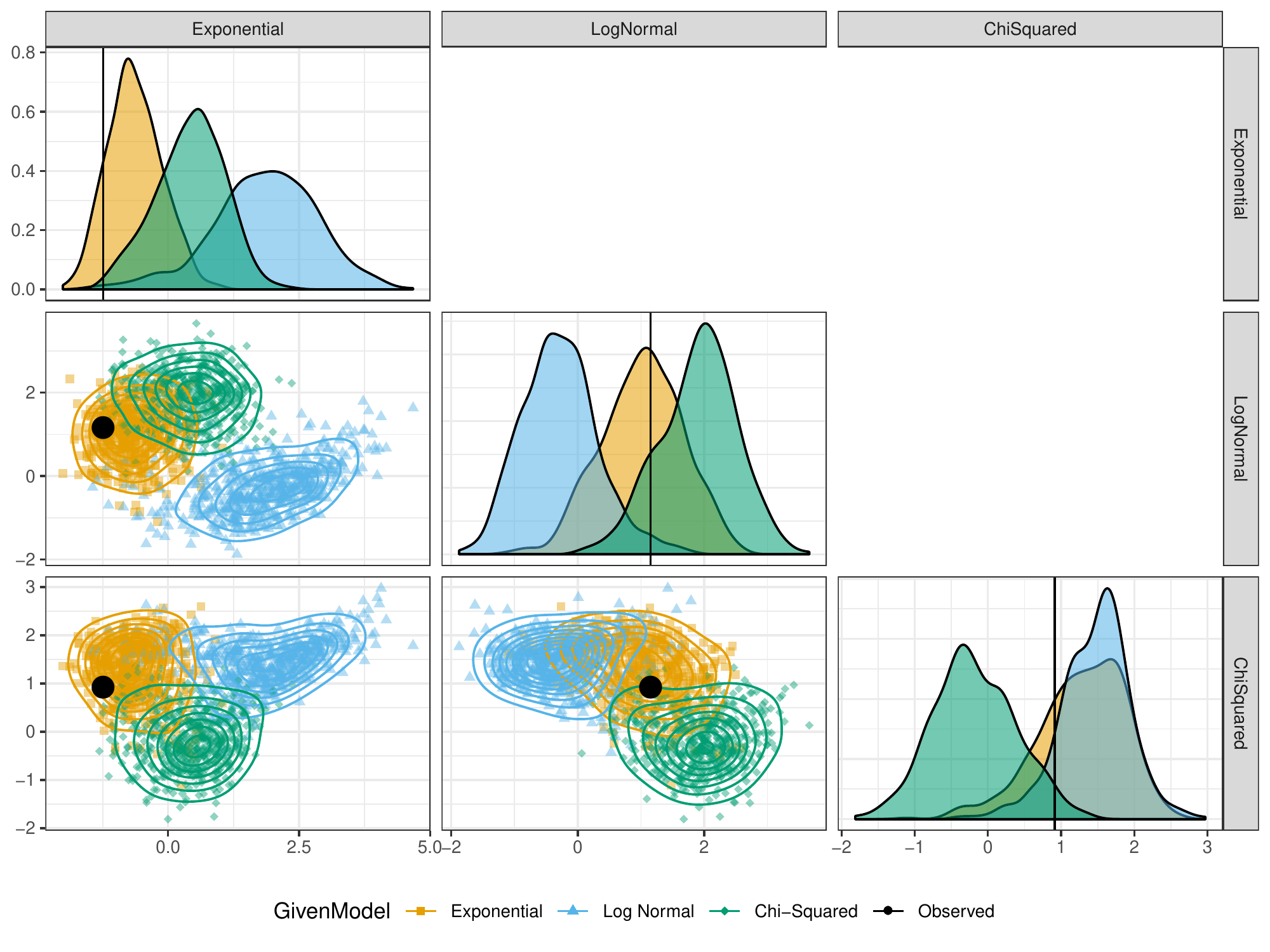} 
		\caption[Example of multi-model mimicry, for simulated exponential data: Pairwise plots]{Plot comparing the fit of exponential, log-normal and chi-squared distributions to data simulated from the exponential distribution, using multi-model mimicry. Each plot on the diagonal is of marginal, univariate densities of energy goodness-of-fit statistics to the labelled model, of data simulated from the three models as denoted in the legend. For example, the plot in the second row and second column is of how well data simulated under all three models fits to the log-normal distribution; the black line in this plot is the observed goodness-of-fit to the log-normal distribution. Scatterplots below the diagonal are of goodness-of-fit statistics to the two labelled models, of data simulated from the three models as denoted in the legend. For example, the plot in the second row and first column is of goodness-of-fit to the exponential (\(x\)-axis) and log-normal distributions (\(y\)-axis), of data simulated under all three distributions. A contour plot for each hypothesis is added to aid interpretation. The black dot in this scatterplot corresponds to the observed goodness-of-fit. The exponential distribution appears to be the best fit to the data.}
		\label{fig:PBCMegpairs}
	\end{center}
\end{figure}

For a visual representation of this, pairwise plots of logarithms of two components of each \(\bm{GOF}\) are given in Figure \ref{fig:PBCMegpairs}, alongside marginal distributions of logarithms of each component of each \(\bm{GOF}\).
Logarithms are taken to make the plots more easy to visualise. Pairwise plots are used due to the difficulty of representing the three-dimensional clouds of variates from the vectors of goodnesses-of-fit. Each of \(\bm{GOF_{exp}},\) \(\bm{GOF_{logN}}\) and \(\bm{GOF_{\chi^2}}\) are represented by a cluster of points in the pairwise scatterplots, and by a distribution function on the diagonals of this figure to represent the marginal distributions under each model. The observed value, \(\bm{GOF_{obs}}\), is represented by a black point or black line. It can be seen that the observed value appears to fit most closely to the exponential variates in all plots, indicating that the observed goodness-of-fit is more likely given that the true model is the exponential distribution, than if the true model is log-normal or chi-squared. Thus, the exponential model should be selected.

However, it is not sufficient to show that components of \(\bm{GOF}_{obs}\) are, marginally, closest to components of the variates from \(\bm{GOF}_{exp}\). The components of \(\bm{GOF}_{obs}\) are marginal distributions of the goodness-of-fit, and marginal distributions do not completely characterise a joint distribution. To confirm that the exponential model should be selected, linear, quadratic and mixture discriminant analyses were undertaken to classify the point \(\bm{GOF}_{obs}\). All discriminant analyses classified the observed goodness-of-fit into the exponential cluster, with probabilities 0.665, 0.998, and 0.997 for linear, quadratic and mixture discriminant analyses respectively.

\begin{figure}
	\begin{center}
		\includegraphics[width=0.5\textwidth]{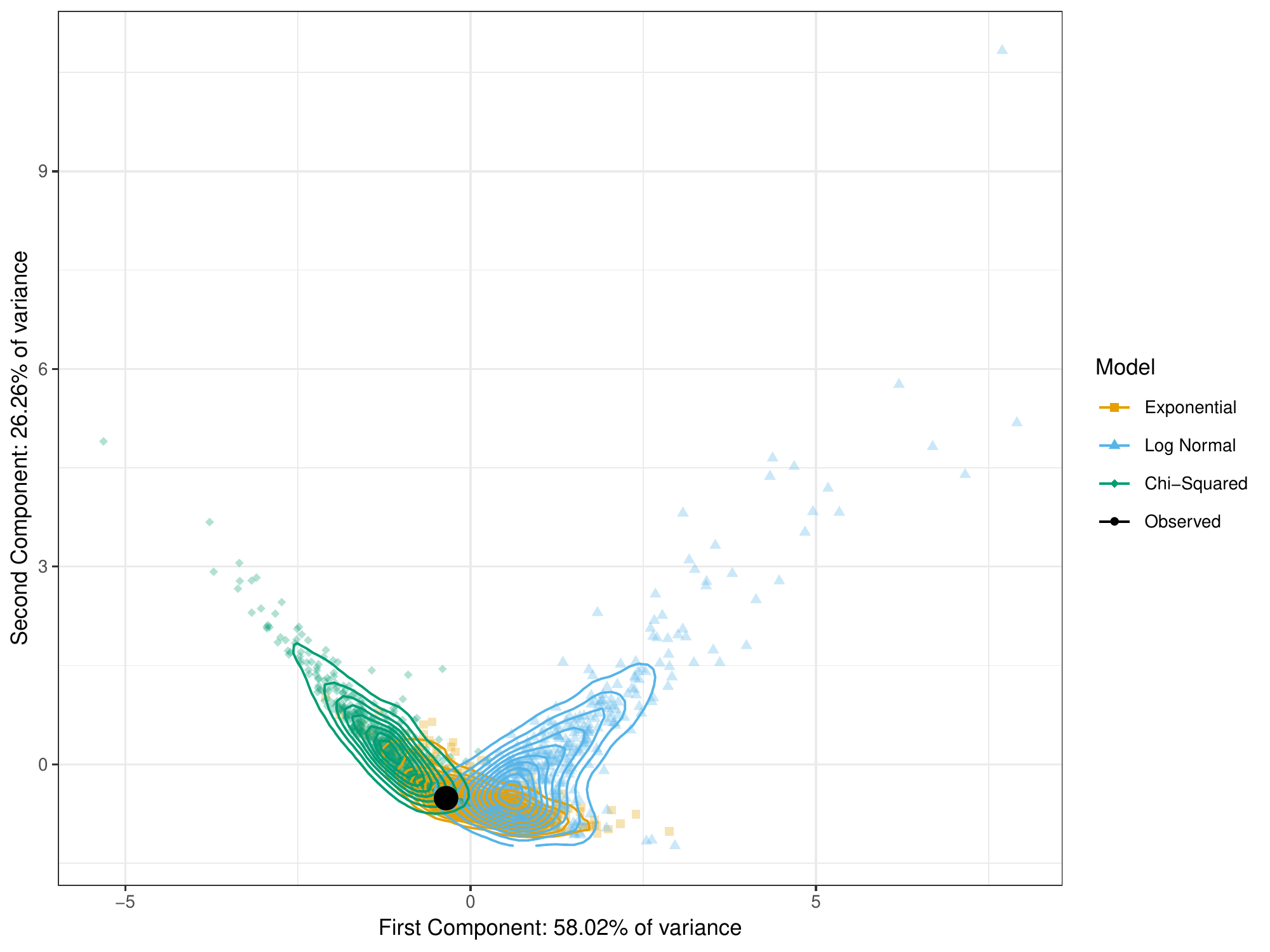} 
		\caption[Example of multi-model mimicry, for simulated exponential data: Principal components plot]{Plot comparing principal components of the fit of the exponential, log-normal and chi-squared distributions to data simulated from the exponential distribution, using multi-model mimicry. Each cluster of points represents data simulated under one of the hypotheses: that the data comes from an exponential (orange), a log-normal (blue) or a chi-squared distribution (green). A contour plot for each hypothesis is added to aid interpretation. Since the black point, the observed goodness-of-fit, fits in the centre of the exponential cluster, the exponential model is to be preferred.}
		\label{fig:PBCMegPCA}
	\end{center}
\end{figure}

Further visual representation can be provided by taking principal components of the collection of variates from each of \(\bm{GOF_{exp}},\) \(\bm{GOF_{logN}}\) and \(\bm{GOF_{\chi^2}}\), and projecting \(\bm{GOF_{obs}}\) onto this space, providing two-dimensional projections of the goodness-of-fit space. A plot of this can be found in Figure \ref{fig:PBCMegPCA} and reinforces the selection of the exponential model. In this instance, the first two principal components capture 84.3\% of the variance in the \(\bm{GOF}\) distributions, meaning Figure \ref{fig:PBCMegPCA} is sufficiently characteristic of the data to be of use in model selection.

In summary, multi-model mimicry allows for the comparison of multiple models simultaneously, using any relevant goodness-of-fit measure. The approach is effective at distinguishing between models even at small sample sizes; here \(n=100\). While the final act of model classification is difficult to directly visualise, pairwise scatterplots and principal components plots can aid this. Classification can then be undertaken by inspection or using discriminant analyses, with discriminant analyses providing a clearer understanding of the level of uncertainty in model choice.

\section{Discussion of the multi-model mimicry framework} \label{sec:appext}
The multi-model mimicry technique introduced by Schultheis and Naidu, and given broader statistical underpinnings in this manuscript, is useful for comparing any number of candidate models with the following capabilities:
\begin{enumerate}
    \item Model parameters should be estimable given training data.
    \item There must exist some statistic by which the data's goodness-of-fit to the model can be measured.
    \item It must be possible to simulate data under the model.
\end{enumerate}
This makes MMM more flexible than a number of alternative techniques for model comparison. For example, relative to Wilks' likelihood-ratio test, MMM does not have the requirement that the models need be nested. Relative to direct comparisons of AIC or BIC, MMM does not have the requirement that likelihood can be computed and that the models and data are of a form that allow AIC and BIC values to directly be compared. Relative to Approximate Bayesian Computation, MMM does not require the estimation of prior probabilities of models and their parameters, and can be undertaken in a frequentist environment.

That MMM is more flexible than some alternatives does not, however, make it universally applicable; the three requirements above are non-trivial. These requirements will now be discussed in turn, followed by a fourth limitation--the MMM's preference for complexity.

\paragraph{Parameter estimation}
The MMM technique requires the model to be able to be fit, based upon the observed data. The technique does not provide alternative methods for such model fitting, so if a model is poorly specified or unidentifiable, the MMM procedure does not resolve this issue. Computational approaches such as Markov Chain Monte Carlo or Approximate Bayesian Computation may resolve these issues in some cases.

\paragraph{Goodness-of-fit statistics}
The MMM technique requires goodness-of-fit statistics for each model to be defined, and calculable. It is assumed in this manuscript that the same goodness-of-fit measure is used for all candidate models, but it may be the case that the MMM procedure is still useful in cases where different goodness-of-fit measures are available for different models. An earlier example given in this manuscript is the difficulty of calculating AIC or BIC for a model incorporating kernel density estimates, since the number of parameters is difficult to determine. Other distributional goodness-of-fit measures can determine how well a kernel density estimate fits to data. It remains uncertain whether within the MMM framework, the distributional goodness-of-fit measures for a kernel density estimate could be compared to AIC or BIC for other candidate models. There is no computational reason that this comparison could not be done, but the statistical consequences of this are a potential course of further research.

\paragraph{Simulation}
The MMM approach involves simulation of data under all candidate models, using the parameters of these models. This naturally brings about questions as to the applicability of the approach to non-parametric models, or those where the confines of the model, and its ability to simulate new data, is harder to define. For example, in many machine learning contexts, such as neural networks, goodness-of-fit to models can be calculated in the form of loss functions, but the simulation of new data under models is an emerging area of research. For example, a deep learning environment such as that in image processing may be able to classify images into certain categories, and develop a model for doing so, but it may be unable to produce new images according to these classifications. Where simulation is not possible, the MMM technique is unavailable. An course of further research may involve the use of non-parametric bootstraps under each model, rather than parametric bootstraps as in the MMM's cross-fitting method, to compare model performance.

\paragraph{Complexity}
When introducing the model mimicry technique for comparison of two models, Wagenmakers \emph{et al.} \citep{Wagenmakers:2004aa} note that the technique has a preference for more complex models, since a more complex model may be able to produce data which mimics that produced by a simpler model. This issue recurs in the extension to multiple models introduced by Schultheis and Naidu \citep{Schultheis:2014aa} and enumerated here. Further research may illuminate methods for penalising complex models in the MMM context. As a general rule for application of the MMM, this issue is addressed by preferring a more parsimonious model over a model complex one, where MMM does not express a strong preference for the more complex model.

\section{Conclusion}
This manuscript presents the history behind, and statistical context for, the multi-model mimicry technique. MMM provides a framework for the simultaneous comparison of multiple statistical models, on the basis of generalised goodness-of-fit measures and in cases where candidate models differ greatly in model or data structure. In doing so, the technique is not limited to scenarios wherein likelihood for the candidate models is calculable, where likelihood-based goodness-of-fit measures are able to be compared between models, or where it is even likelihood that is sought to be compared between models. For example, the MMM technique allows for the comparison of models on the basis of goodness-of-fit to model-building assumptions, to desired features like the frequency or quantification of extreme data points, or to some distribution.

This manuscript also demonstrates the effectiveness of this technique, even at relatively small sample sizes and for otherwise difficult-to-distinguish models. For example, for 100 data points simulated from an exponential distribution, MMM was able to correctly identify that this data came from an exponential distribution, rather than a log-normal or a chi-squared distribution.

Finally, we discuss potential areas for further research in the area of the MMM. How the MMM framework interacts with situations in which model parameters are difficult to determine, where different goodness-of-fit measures are used for different models, or where simulation of new data under models is difficult, all form courses for future exploration. A more sophisticated approach for dealing with MMM's preference towards more complex models would also be a welcome addition to literature in this area.

\section*{Acknowledgements}\label{acknowledgements}

The authors wish to acknowledge Prof. Nigel Bean and Dr. Jonathan Tuke for guidance and assistance in completing this work. Thank also to the ARC Centre of Excellence for Mathematical and Statistical Frontiers (ACEMS) for supporting the completion of this work.

\pagebreak
 \bibliography{bibliography}

\end{document}